\documentclass[twocolumn,showpacs,preprintnumbers,amsmath,amssymb]{revtex4}

\usepackage{graphicx}
\usepackage{dcolumn}
\usepackage{bm}

\begin{document}


\title{Recent BES measurements and the hadronic contribution to the QED vacuum polarization}

\author{H.~Burkhardt}
\email{Helmut.Burkhardt@cern.ch}
\affiliation{%
CERN, CH-1211 Geneva 23,  Switzerland
}%

\author{B.~Pietrzyk}
\affiliation{ Laboratoire de Physique des Particules LAPP,
Universit\'{e} de Savoie, IN2P3-CNRS, F-74941 Annecy-le-Vieux
Cedex, France
}%
\email{pietrzyk@lapp.in2p3.fr}
             \date{June 15, 2011}

\begin{abstract}
We have updated our evaluation of the hadronic contribution to the
running of the QED fine structure constant using the recent
precise measurements of the e$^+$e$^-$ annihilation at the
center-of-mass (c.m.s.) energy region between 2.6 and 3.65\,GeV
performed by the BES collaboration. In the low energy region,
around the $\rho$ resonance, we include the recent measurements
from the BABAR, CDM-2, KLOE and SND collaborations. We obtain
$\Delta\alpha^{(5)}_{\rm had}(s) = 0.02750 \pm 0.00033$ at $s$ =
$m {\rm _Z^2}$.
\end{abstract}

\maketitle

We have been following up on the cross section measurements of
e$^+$e$^-$ annihilation into hadrons for many years, and provide
up-to-date evaluations of the hadronic contribution to the running
of the QED fine structure constant $\Delta\alpha^{(5)}_{\rm
had}(s)$ at $s$~=~$m \rm{_Z^2}$, based on a dispersion integration
which uses the experimental data as
input\,\cite{BurPie95}\cite{BurPie2001}\cite{BurPie2005}.

The measured hadronic cross sections are conveniently given as R$_{\rm had}$, i.e. in units of the QED cross-section for lepton-pair production.

\begin{figure}
\includegraphics[width=8.3cm]{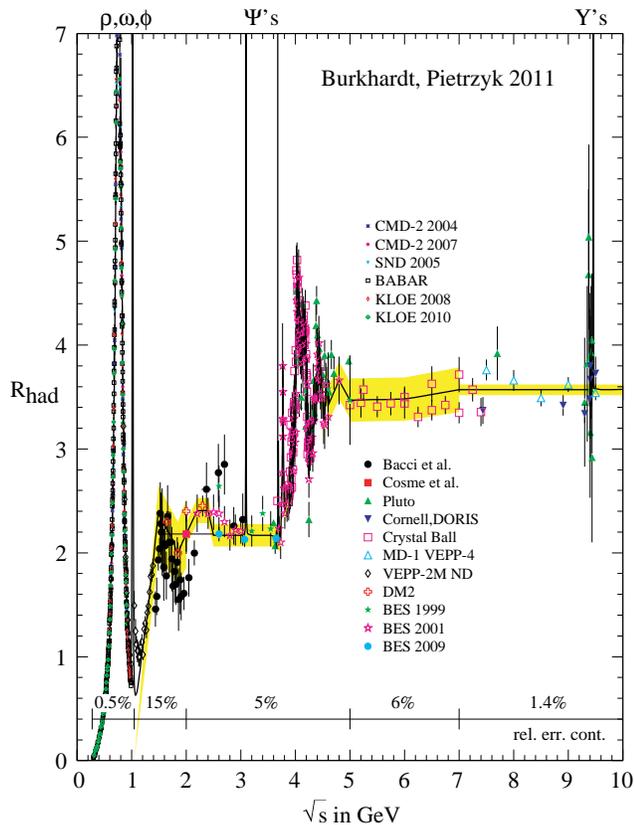}
\caption{\label{plot:RHAD} R$_{\rm had}$ including resonances.
Measurements are shown with statistical errors. The relative
uncertainty assigned to our parametrization is shown as band and
given with numbers at the bottom.} \label{fig:BuPieAlpha11_fig1}
\end{figure}

The BES collaboration has previously measured the R$_{\rm had}$
value for e$^+$e$^-$ annihilation in the c.m.s. energy range
between 2 and 5 GeV \cite{BES2000}\cite{BES2002} (BES1999, BES2001
in Fig.\,\ref{fig:BuPieAlpha11_fig1}). This energy region is
particularly important for the analysis of
$\Delta\alpha^{(5)}_{\rm had}(s)$ \cite{BurPie2001}. These results
were presented at the ICHEP 2000 Conference in Osaka
\cite{ZhaoOsaka} and were included in the evaluation of
$\Delta\alpha^{(5)}_{\rm had}(s)$ \cite{BurPie2001}. This had a
significant impact on the LEP ElectroWeak Group Standard Model
precision measurement fits: the most probable value of the Higgs
mass moved up from 60 GeV to 88 GeV \cite{PieOsaka} giving a more
coherent picture of the Standard Model of particle physics. The
large number of points measured by the BES collaboration, allows
to connect and integrate these directly, taking into account the
correlation between the systematic uncertainties and uncorrelated
statistical errors. As a result, the dispersion integral in the 2
to 5 GeV range was obtained with a precision of 5.9\%.

More recently, the BES collaboration has published \cite{BES2009}
(referred to as BES2009) measurements of R$_{\rm had}$ at 2.60,
3.07 and 3.65\,GeV with statistical errors below the 1\% level,
and systematic errors of about 3.5\%. In order to properly include
these new measurements in our analysis, we have divided the BES1999
and BES2001 data points into three c.m.s. energy regions: the
region covered by the recent BES2009 measurements to which we
refer to as the overlap region, and the regions below and above.
We evaluate the dispersion integral based on the earlier data
separately in these three regions, assuming conservatively that
the systematic errors in these three regions are fully correlated.
The recent BES2009 data gives us an additional, more precise
result for the dispersion integral in the overlap region. This
result is combined with the previous results assuming a
conservative value of 0.5 for the correlation between the
systematic uncertainties in the previous and recent BES data.
We obtain the total uncertainty of the dispersion integral in the
three regions of 7.6\% ("below"), 3.7\% ("overlap") and 5.0\%
("above"). As a result of the inclusion of the recent BES data,
the value of the dispersion integration in the c.m.s. energy
region from 2 to 5 GeV decreases from 0.00381 to 0.00371 and the
overall uncertainty from 5.9\% to 5.0\%.

At very low energies around the $\rho$, we used in our previous publication \cite{BurPie2005},
the results from the CMD-2 collaboration with cross section measurements
in the c.m.s. energy region between 0.61 and 0.96\,GeV
\cite{CMD2-2001} (CMD-2 2004) and the KLOE collaboration pion form
factor data using the "radiative return" from the $\phi$ resonance
to the $\rho$ in the $\pi^+\pi^-$ mass range between 0.59 and
0.97\,GeV \cite{KLOE2004}. The small $\rho$ contribution from
lower and higher energies, not covered by data, was evaluated
using the CMD-2 parametrization of the pion form factor
\cite{CMD2-2001}.

In the low energy region, around the $\rho$ resonance, we now
include in our analysis all recent measurements from the BABAR,
CDM-2, KLOE and SND collaborations.
The KLOE collaboration has superseded the previous measurements by
new ones \cite{KLOE2008} (KLOE2008) in the same c.m.s. energy
region. In addition a new analysis \cite{KLOE2010} (KLOE2010), in
which the "radiative return" photon was detected in the detector,
has allowed the collaboration to extend the $\pi^+\pi^-$ mass
range down to the threshold for the di-pion production. The CMD-2
\cite{CMD2-2007} (CMD-2 2007) and SND collaborations
\cite{SND2006} (SND 2005) at Novosibirsk have published new
measurements in the c.m.s. energy region between 0.6 and 0.97\,GeV
and between 0.39 and 0.97\,GeV, respectively. Finally, the BABAR
collaboration \cite{BABAR2009} has used the "radiative return"
measurements from the c.m.s. energies near 10.6\,GeV to measure
the $\pi^+\pi^-$ cross section from threshold up to a c.m.s.
energy of 3\,GeV.

The contribution of the new results on $\Delta\alpha^{(5)}_{\rm
had}(m {\rm _Z^2})$ was obtained by direct integration between
measured data points and  the small $\rho$ contribution from lower
and higher energies, not covered by data, was evaluated using the
CMD-2 parametrization of the pion form factor \cite{CMD2-2001} as
in our previous analysis \cite{BurPie2005}. We found excellent
agreement between the dispersion integral results of all these measurements.
We have combined them assuming full correlation between
systematic uncertainties in the same experiment, CMD-2 and KLOE,
and no correlation between different experiments. The value of the
$\rho$ dispersion integration has increased from 0.00347 in
\cite{BurPie2005} to 0.00349 and the relative uncertainty has
decreased from 0.9\% to 0.5\%.

Fig.~\ref{plot:RHAD} and Table~\ref{tab:contrib} give the
summary of R$_{\rm had}$ measurements by different experiments and
the current precision in different e$^+$e$^-$ c.m.s. energy
regions.

\begin{table}[h]
\caption{\label{tab:contrib}Contributions to
$\Delta\alpha^{(5)}_{\rm had}(m {\rm ^2_Z})$}
\begin{ruledtabular}
\begin{tabular}{ccr}
Range $\sqrt{s}$, GeV& $\Delta\alpha$ & Relative error\\ \hline
$\rho$                 & 0.00349 & 0.5 \% \\
Narrow resonances      & 0.00184 & 3.1 \% \\
1.05 -- 2.0\phantom{0} & 0.00156 & 15 \% \\
2.0 -- 5.0             & 0.00371 & 5.0 \%  \\
5 -- 7                 & 0.00183 & \phantom{0}6 \%  \\
\phantom{0}7 -- 12     & 0.00304 & 1.4 \% \\
\phantom{00}$>$ 12     & 0.01203 & 0.2 \%  \\
\hline
 & 0.02750 & 1.2 \% \\ \hline
\end{tabular}
\end{ruledtabular}
\end{table}

We obtain a value of the hadronic contribution to the running of
the QED fine structure constant of $\Delta\alpha^{(5)}_{\rm
had}(s)$ = 0.02750 $\pm$ 0.00033 at $s$ = $m {\rm _Z^2}$
corresponding to $1/\alpha^{(5)}(m {\rm _Z^2}) = 128.951 \pm
0.045$. Similar values for $\Delta\alpha^{(5)}_{\rm had}(m {\rm
_Z^2})$ have recently been obtained by M. Davier et al. of 0.02749
$\pm$ 0.00010 \cite{Davier2010} and by the HLMNT group
\cite{HLMNT2010} of 0.02760 $\pm$ 0.00015. The uncertainty quoted
by us for the running of the fine structure constant is directly
obtained from the experimental uncertainties. The uncertainties
obtained in Refs.\,\cite{Davier2010} and \cite{HLMNT2010} are
reduced by relying on perturbative QCD to calculate R$_{\rm had}$
in various regions, including the region of the recent
measurements of the BES collaboration.

A simple parametrization of the hadronic contribution to the
vacuum polarization as a function of energy
\cite{TASSONOTE,HADVACPO} is used in many computer programs. This
parametrization provides a description of the dispersion integral
result within 0.2 $\sigma$ in the whole $t$-channel and the exact
value at $m {\rm _Z}$ in the $s$-channel. The computer code for
$\Delta\alpha^{(5)}_{\rm had}(s)$ with this parametrization is
available from the authors.

Our new value of $\Delta\alpha^{(5)}_{\rm had}(m {\rm _Z^2})$ =
0.02750 $\pm$ 0.00033 increases the preferred Higgs mass value
from 89$^{+35}_{-26}$ to 93$^{+35}_{-27}$\,GeV
 and the one-sided 95\% confidence level upper limit from 158 to 163\,GeV \cite{LEPEWWG}.


\begin{thebibliography}{99}
\bibitem{BurPie95} H. Burkhardt and B. Pietrzyk, Phys. Lett. {\bf B356} (1995) 398.
\bibitem{BurPie2001} H.~Burkhardt and B.~Pietrzyk, Phys. Lett. {\bf B513} (2001)46.
\bibitem{BurPie2005} H. Burkhardt and B. Pietrzyk, Phys. Rev. {\bf D72}:057501, 2005.
\bibitem{BES2000} J.Z.~Bai et al., BES Collaboration, Phys. Rev. Lett.
{\bf84} (2000) 594.
\bibitem{BES2002} J.Z. Bai et al., BES Collaboration, Phys. Rev.
Lett. {\bf 88}:101802, 2002.
\bibitem{ZhaoOsaka} Z.G. Zhao, Proc. of ICHEP 2000, Osaka,
Japan, 27 July-2 August 2000, Ed. C.S. Lim and T. Yamanaka, page
644.
\bibitem{PieOsaka} B. Pietrzyk, Proc. of ICHEP 2000, Osaka, Japan, 27 July-2 August 2000,
Ed. C.S. Lim and T. Yamanaka, page 710.
\bibitem{BES2009} M. Ablikim et al., BES Collaboration, Phys. Lett. {\bf B677} (2009) 239.
\bibitem{CMD2-2001} R.R.~Akhmetshin et al., CMD-2 Collaboration, hep-ex/9904027, Phys.  Lett. {\bf B527}
 (2002)161; {\bf B578} (2004)285.
 \bibitem{KLOE2004} A.~Aloisio et al., KLOE Collaboration, Phys.  Lett. {\bf B606} (2005)12.
 \bibitem{KLOE2008} F. Ambrosino et al., KLOE Collaboration,  Phys. Lett. {\bf B670} (2009)
 285.
\bibitem{KLOE2010} F. Ambrosino et al., KLOE Collaboration,
arXiv:hep-ex/1006.5313.
\bibitem{CMD2-2007} R.R. Akhmetshin et al., CMD-2 Collaboration,
Phys.Lett. {\bf B648} (2007) 28.
\bibitem{SND2006} M.N. Achasov et al., SND Collaboration,
J.Exp.Theor.Phys. {\bf 103} (2006) 380.
\bibitem{BABAR2009} B. Aubert et al., BABAR Collaboration, Phys.
 Rev. Lett. {\bf 103}:2311801, 2009.
\bibitem{Davier2010} M. Davier et al., Eur.Phys.J. {\bf C71} (2011)
1515.
\bibitem{HLMNT2010} T. Teubner et al., arXiv:1001.5401, Chinese
Physics C {\bf 34} (2010) 728.
\bibitem{TASSONOTE} H. Burkhardt, TASSO Note 192 (1981) and DESY F35--82--03
(thesis).
\bibitem{HADVACPO} H. Burkhardt, F. Jegerlehner, G. Penso and C. Verzegnassi,
Z. Phys.  C43  (1989) 497.
\bibitem{LEPEWWG} M. Gruenewald, LEP Electroweak Working Group (http://lepewwg.web.cern.ch/LEPEWWG/), private communication.

\end{thebibliography}
\end{document}